\documentclass[runningheads]{llncs}

\usepackage{amsmath}
\usepackage{amsfonts}
\usepackage{amssymb}
\usepackage{booktabs}
\usepackage{caption}
\usepackage{cite}
\usepackage{subcaption}
\usepackage{multirow}
\usepackage{multicol}
\usepackage{mathtools}
\usepackage{url}
\usepackage{hyperref}
\usepackage[svgnames]{xcolor}
\usepackage{footnote}
\usepackage{epsfig}
\usepackage{longtable}
\usepackage{graphicx}
\usepackage{pgfplots}
\usepackage{array,tabularx}

\usepackage[symbol]{footmisc}

\usepackage{algpseudocode}

\usepackage{cleveref}

\usepackage[boxruled,longend,linesnumbered,ruled]{algorithm2e}
\hypersetup{
     colorlinks = true,
     linkcolor = blue,
     citecolor = red,
     filecolor = blue,
     urlcolor = blue
     }	
\usepackage{bm}
\usepackage{tikz}
\usetikzlibrary{shapes.geometric, arrows}
\tikzstyle{startstop} = [rectangle, rounded corners, minimum width=3cm, minimum height=1cm,text centered, draw=black, fill=white!30]
\tikzstyle{io} = [trapezium, trapezium left angle=70, trapezium right angle=110, minimum width=3cm, minimum height=1cm, text centered, draw=black, fill=white!30]
\tikzstyle{process} = [rectangle, minimum width=3cm, minimum height=1cm, text centered, draw=black, fill=white!30]
\tikzstyle{decision} = [diamond, minimum width=3cm, minimum height=1cm, text centered, draw=black, fill=white!30]
\tikzstyle{arrow} = [thick,->,>=stealth]

\begin{document}

\title{ Towards an Adaptive Dynamic Mode Decomposition  \thanks{This work is funded by the Office of Research, North South University, under grant number \textbf{CTRG-19/SEPS/06}
}}
\titlerunning{}
\author{Mohammad N. Murshed \inst{1}\orcidID{0000-0001-9859-2523} \and 
M. Monir Uddin\inst{4}\orcidID{0000-0002-9817-6156}}
\authorrunning{M. N. Murshed \textit{et al.}}
\institute{Department of Mathematics and Physics, North South University, Dhaka-1229, Bangladesh, \email{mohammad.murshed@northsouth.edu} \\  \and Department of Mathematics and Physics, North South University, Dhaka-1229, Bangladesh, \email{monir.uddin@northsouth.edu}}
\maketitle        
\begin{abstract}
Dynamic Mode Decomposition (DMD) is a data based modeling tool that identifies a matrix to map a quantity at some time instant to the same quantity in future. We design a new version which we call Adaptive Dynamic Mode Decomposition (ADMD) that utilizes time delay coordinates, projection methods and filters as per the nature of the data to create a model for the available problem. Filters are very effective in reducing the rank of high-dimensional dataset. We have incorporated 'discrete Fourier transform' and 'augmented lagrangian multiplier' as filters in our method. The proposed ADMD is tested on several datasets of varying complexities and its performance appears to be promising. 

%In our proposal, we discussed about two projection matrices: one is inspired by the Krylov subspace and the other promotes sparsity to bring computational benefits in producing a model. Satisfactory results are obtained as they are tested on data related to Double gyre (present in ocean mixing) and on a 2D compressible signal. The motivation behind this scheme of DMD comes from the fact that data from many phenomena are 'big' and 'highly oscillatory.'


\end{abstract}
\section{Introduction}
Dynamic Mode Decomposition, \cite{schmid2010dynamic}, is a data-informed modeling technique that has been in use since 2007. If we have some spatiotemporal data available for a phenomenon from any field like fluid dynamics, stock market or epidemiology, then DMD can extract the dynamics from the data and predict the states in the future. This equation free method is able to construct an approximate model for most of the problems, but need to be updated for a few of them. For instance, DMD would have to use time delay coordinates in case the data is highly oscillatory, \cite{9038561}. There are many versions of DMD designed to suit different kinds of problems.\\ \\
Often, the data is high dimensional and processing all of them in the DMD algorithm may not be a good idea. The term 'high-dimensional' can refer to two different situations -- 'more snapshots than pixels' and 'more pixels than snapshots.' \\ \\
\textbf{More snapshots than pixels.} Cases where the number of snapshots is higher than the number of pixels in a snapshot are well addressed by Streaming Dynamic Mode Decomposition, \cite{hemati2014dynamic}, and Online Dynamic Mode Decomposition, \cite{zhang2019online}.\\ \\ 
\textbf{More pixels than snapshots.} This implies that the state dimension is higher than the time dimension in the data-set. Compressing such high dimensional data to lower dimensional space yields a matrix that will save both memory and time as DMD processes to come up with a model. This idea of compression is also known as Compressed DMD, \cite{brunton2013compressive}, by the practitioners. \\ \\
Sampling and projections have been used as a pre-processing step in DMD to minimize computational burden. The data sequence to be used as the input into the DMD algorithm is represented by a small matrix known as sketch. Illustrated in \cite{kutz2016dynamic,murshed2020projection} is how compressive sampling theories can be applied to the input matrix to get a compressed version of the available data to efficiently produce a model. Tu proposed a compressed DMD routine and got promising results after its application on a compressible signal, \cite{jtu}. Authors in \cite{erichson2019randomized} deployed the tool on sea surface temperature (SST) data. Randomized DMD associates sampling/projection methods with the idea of DMD to develop reduced order models for complex fluid flows, \cite{bistrian2017randomized}. \\ \\
Unlike compressible signals, data in unsteady flows exhibit behavior at multiple scales. For example, it is quite difficult to find the pattern of the changes in a particular flow variable in a turbulent channel flow even after we collect data for several periods of oscillation. Furthermore, such dataset contains noise/outliers that can reduce the quality of the DMD model. Research is under way on how to update modeling tools to deal with highly non-linear flows in fluid mechanics. \\ \\
We provide, in this work, an architecture of DMD that will operate based on the nature of dataset. The rank of the dataset plays an integral role in determining the complexity of the data, hence, this parameter is to be used to recommend the appropriate pre-processing step prior to the dynamic mode decomposition of the data. Any low-rank dataset will be processed by delay coordinates and projection methods to extract the dominant and the hidden features at a low cost. The user can choose between Gaussian projection and sparse random projection depending on the need. For high-rank data, either discrete Fourier transform or inexact augmented Lagrange multiplier method is to be used as per the level of noise present in the data. \\ \\
The rest of the paper is organized as follows: Section \ref{bg} is a discussion of the basics of DMD, projection methods and delay coordinates that will be used in the making of the adaptive DMD, elaborated in Section \ref{maco}. The numerical results are presented in Section \ref{numres} and a summary provided in Section \ref{cfw}. 

\section{Background}
\label{bg}
 This section presents the theories, definitions and algorithms necessary to develop the Adaptive DMD. The reader may take a look at \cite{trefethen1997numerical} to review some of the basic concepts of linear algebra (if necessary). In particular, we discuss the  basics of dynamic mode decomposition: time delay coordinates and projection methods. 

\subsection{Dynamic Mode Decomposition}
Let's assume that we have some spatiotemporal data, \(\textbf{X} \in \mathbb{R}^{M \times N}\),
\[\textbf{X}=\begin{bmatrix}
    \textbf{x}_{11} & \textbf{x}_{12} &  ... & \textbf{x}_{1N} \\
    \textbf{x}_{21} & \textbf{x}_{22} &  ... & \textbf{x}_{2N} \\
    \textbf{x}_{31} & \textbf{x}_{32} &... & \textbf{x}_{3N} \\
     \vdots            & \vdots              &  \ddots & \textbf{x}_{(M-1)N} \\
    \textbf{x}_{M1} & \textbf{x}_{M2} &  ... & \textbf{x}_{MN} \\
    \end{bmatrix},
\]
for a problem in the form of a matrix where the columns represent a certain quantity for different spatial coordinates at a particular time instant. \(M\) is the number of spatial nodes and \(N\) the number of temporal nodes. Dynamic Mode Decomposition splits this data sequence into two parts, \(\textbf{X}_1^{N-1}\) and \(\textbf{X}_2^{N}\), and runs according to Algorithm \ref{ALGO:1}, \cite{jtu}, to generate a model to predict the future states. A critical aspect of DMD is the sampling frequency used in obtaining the data. For DMD to work properly, data must be sampled at exactly equal to or greater than twice the maximum frequency in the data. In some exceptional cases when data acquisition is difficult, DMD can be improved to identify modes from sub-Nyquist-rate data, \cite{tu2014spectral}.\\

    \begin{algorithm}
    \caption{DMD}\label{ALGO:1}
     \SetAlgoLined
      \KwData{\(\textbf{X}_{1}^{N-1}, \textbf{X}_{2}^{N}\)}
      \KwResult{$\textbf{x}_{DMD}(t)$}
    $ \textbf{X}_1^{N-1}=\textbf{U}\Sigma \textbf{V}^{*}$\\
    $ r \ = \ rank(\textbf{X}_{1}^{N-1}) $\\
    $ \textbf{U}=\textbf{U} [ \ :  \ , 1  : r ] $\\
    $ \Sigma= \Sigma [ 1 :  r , 1 : r ] $\\
    $ \textbf{V}=\textbf{V} [ \ :  \ , 1 : r ] $\\
    $\tilde{S} = \textbf{U}^{*}\textbf{X}_2^{N} \textbf{V} \Sigma^{-1}$\\
    $\tilde{S} \textbf{y}_{k} = \mu_{k} \textbf{y}_k$\\
    $\phi_{k} = \textbf{U} \textbf{y}_{k}$ \\
    $\omega_{k}= \frac{ln(\mu_{k})} {\Delta t} \) and \(\textbf{b}=\Phi^{\dagger}\textbf{x}_1$\\
    $\textbf{x}_{DMD}(t) =  \sum_{k=1}^{r} b_{k}(0) \phi_{k}(\textbf{x}) exp(\omega t) = \Phi \ \ diag(exp(\omega t)) \textbf{b} $
    \end{algorithm}

\noindent \textbf{Connection to Koopman Operator and Proper Orthogonal Decomposition}\\ \\
DMD is a special case of Koopman Mode Decomposition (KMD), \cite{kutz2016dynamic}. The key difference is that DMD works on state space whereas KMD takes place in observable space:
$$ g(\textbf{x}_{k+1})=\textbf{K}g(\textbf{x}_k). $$ 
An interesting aspect in DMD is the singular value decomposition of the first data sequence, \(\textbf{X}_1^{N-1}\), that results in \(\textbf{U}\). The columns of \(\textbf{U}\) are also known as the modes from the proper orthogonal decomposition (POD),  \cite{sirovich1987turbulence, chatterjee2000introduction}. POD searches for as few (optimal) orthogonal bases as possible to approximate the flow. Next, we reveal the critical points of DMD and a few situations where DMD fails to learn from data.

\subsection{Time delay coordinate based DMD}
Time delay coordinate based DMD, \cite{kutz2016dynamic}, is a variant of DMD that uses augmented data sequence (in the form of Hankel matrices) to generate a model. The dataset is augmented vertically by adding a copy of the time shifted states. Consider the state vectors for 6 temporal nodes 
\(\textbf{x}_{1}, \textbf{x}_{2}, \textbf{x}_{3}, 
\textbf{x}_{4}, \textbf{x}_{5}, \textbf{x}_{6}\). We can 
construct \(\textbf{X}_{1, aug}\) and \(\textbf{X}_{2, aug}\) (before running DMD) as
\[
\textbf{X}_{1, aug}=\begin{bmatrix}
    \textbf{x}_{1} &  \textbf{x}_{2} &  \cdots &  \textbf{x}_{N-q}\\
    \vdots & \vdots  & \vdots & \vdots\\
    \textbf{x}_{q}  &  \textbf{x}_{q+1} & \cdots & \textbf{x}_{N-q+1}\\
    \end{bmatrix}, \ 
\textbf{X}_{2, aug}=\begin{bmatrix}
    \textbf{x}_{2} &  \textbf{x}_{3} &  \cdots &  \textbf{x}_{N-q+1}\\
    \vdots & \vdots  & \vdots & \vdots\\
    \textbf{x}_{q+1}  &  \textbf{x}_{q+2} & \cdots & \textbf{x}_{N}\\
    \end{bmatrix}.
\]

The routine for time delay coordinate based DMD is provided in Algorithm \ref{ALGO:2}. We define the variable \(q\) as 
$$ q =  1 +  p, $$
where \(p\) is the number of time shifted states. For the instance above, 1 copy of time shifted data is used, hence, \(q = 2\). It is imperative to carefully set the value of \(q\) so to capture the dynamic modes.  This algorithm is essentially classic DMD when \(q=1\). \\

    \begin{algorithm}
    \SetAlgoLined
      \KwData{\(\textbf{X}_{1,aug}^{N-1}, \textbf{X}_{2,aug}^{N}\)}
      \KwResult{\(\textbf{x}_{DMD}(t)\)}
    \caption{Time Delay Coordinate based DMD}\label{ALGO:2}
    $ \textbf{X}_{1,aug}^{N-1}=\textbf{U}\Sigma  \textbf{V}^{*} $\\
    $ r \ = \ rank(\textbf{X}_{1,aug}^{N-1}) $\\
    $ \textbf{U}=\textbf{U} [ \ :  \ , 1 : r ] $\\
    $ \Sigma= \Sigma [ 1 :  r , 1 : r ] $\\
    $ \textbf{V}=\textbf{V} [ \ :  \ , 1 : r ] $\\
    $\tilde{S} = \textbf{U}^{*}\textbf{X}_{2,aug}^{N} \textbf{V} \Sigma^{-1}$\\
    $\tilde{S} \textbf{y}_{k} = \mu_{k} \textbf{y}_k$\\
    $\phi_{k} = \textbf{U} \textbf{y}_{k}$\\
    $\omega_{k}= \frac{ln(\mu_{k})} {\Delta t} \) and \(\textbf{b}=\Phi^{\dagger}\textbf{x}_1$\\
    $\textbf{x}_{DMD}(t) =  \sum_{k=1}^{r} b_{k}(0) \phi_{k}(\textbf{x}) exp(\omega t) = \Phi \ \ diag(exp(\omega t)) \textbf{b} $\\
    $\textbf{x}_{DMD}(t) =  \textbf{x}_{DMD}(t) [ 1 :  M, 1 :  N ] $
    \end{algorithm}

\subsection{Projection methods for DMD}
Projection is a technique \cite{bingham2001random} to transform a higher dimensional matrix to a lower dimensional one by the use of a random matrix \((\textbf{R})\). The equation to move into a lower dimensional subspace \((a \ll M)\) will read:
\begin{equation}
\textbf{X}_{a \times N}= \textbf{R}_{a \times M} \textbf{X}_{M \times N}.
\end{equation}
The matrix \(\textbf{R}\) is in general orthogonal. The gram matrix must be or somewhat close to identity matrix i.e.
\begin{equation}
\textbf{R}^{*} \textbf{R} \approx \textbf{I}. 
\end{equation}
Many different random matrices have been previously used with DMD. Uniform/Gaussian projections and single pixel measurement were applied to identify the dynamic modes for sparse linear dynamics in the Fourier domain in\cite{kutz2016dynamic}. In randomized DMD (rDMD) \cite{erichson2019randomized}, a massive dataset gets converted to a smaller matrix via sampling and then post-processed by standard DMD to result in the dominant dynamic modes and eigenvalues. Overall, this method takes into account sampling and power iteration to adjust the accuracy of the model. Arnoldi iteration, \cite{trefethen1997numerical, saad1992numerical}, is an iterative method, based on modified Gram-Schmidt orthogonalisation, that takes a random matrix and an arbitrary vector to generate a matrix containing orthonormal bases and an upper triangular matrix. The resulting orthonormal matrix is then transposed and multiplied by the data sequence to project the high dimensional data to lower dimensional subspace. As a sparse random projector matrix, Achlioptas is a notable one. Achlioptas \cite{achlioptas2001database} proposed a random projection matrix with entries based on the distribution below with \(s\) being either 1 or 3:
\[ R_{ij}=\sqrt{s}\begin{cases} 

     -1 & \text{ with  probability $1/(2s)$}\\

      0 & \text{ with  probability $1-1/s$} \\

      1 & \text{ with  probability $1/(2s)$} 

   \end{cases}.
\]
Such matrix allows processing of just a fraction of the complete high dimensional data, thereby saving memory and time.\\ \\
The large size of the augmented matrices reduces computational efficiency if \(M \gg N\). Random sampling aids in bypassing this problem. Sampling means to collect a subset of the complete signal. There are several different ways in statistics to sample from a population. Here, we deploy random sampling without replacement to identify a subset of the all the spatial nodes in the domain. This is much like placing sensor on a few locations rather than using sensors at all the spatial nodes in the domain. Although, fewer measurements are expected, it is instructive to have enough measurements to retain the structure of the original signal at a given a spatial coordinate.

\section{Adaptive Dynamic Mode Decomposition}
\label{maco}
In this section, we propose a new version of DMD - which we call Adaptive DMD (ADMD). This scheme would, at first, find the rank of the raw data that is available. If the data is of low-rank, then time-delay coordinates are to be used to resolve smooth oscillations and projections would be employed to extract the hidden features. In case the data contains multiple dominant modes, then discrete Fourier transform based filter is to be applied on the raw data to remove the low amplitude high frequency modes in the data. This works fine when the dataset has small amount of the noise. But, we can opt for augmented Lagrange multiplier method \cite{wright2009robust, yuan2009sparse} in case a few entries in the data matrix contain noise of large magnitude.\\ \\
The steps of ADMD are outlined below:\\ \\
(1) Determine the rank of the data matrix, \(r\), based on the rank threshold \(\epsilon = 10^{-12}\). \\ \\
(2) If the data matrix
\begin{itemize}
    \item is rank deficient (\(r \ll min(M,N)\)), then utilize time-delay coordinates and projection methods to pre-process the data. \\
    \item has full rank (\(r \leq min(M,N)\)), then use either discrete Fourier transform based spatial filter to pre-process the data or inexact augmented lagrange multiplier method, \cite{lin2010augmented}, to recover a low-rank version of the data. In the latter, the dataset is decomposed into a lower rank matrix and a sparse matrix,
   
    $$ \textbf{X} = \textbf{X}_{low \ rank} + \textbf{X}_{sparse} $$
    by solving the convex optimization problem:
    \begin{equation}
    \begin{array}{rrclcl}
    \displaystyle \min_{\textbf{X}_{low \ rank},\textbf{X}_{sparse}} & \multicolumn{3}{l}{||\textbf{X}_{low \ rank}||_{*} + \lambda_{0} ||\textbf{X}_{sparse}||_{1}}\\
    \textrm{s.t.} & \textbf{X}=\textbf{X}_{low \ rank}+\textbf{X}_{sparse}\\
    \end{array}
    \end{equation}
    where \(|| \  ||_*\) is the nuclear norm of a matrix (the sum of the singular values in the matrix) and \(|| \ ||_1\) the sum of the absolute values of the entries in the matrix and \(\lambda_0\) a weighing factor. The weighting factor  \(\lambda_0\) is equal to \(\frac{\lambda}{\sqrt{max(M,N)}}\) and experience has shown that the optimal value of \(\lambda\) is 1.
\end{itemize}
(3) Compute DMD eigenvalues and modes in the processed data.\\ \\
% \begin{algorithm}
%         \caption{DFT-based filter, \cite{cochran1967fast}}\label{dftDMD}
%         \For{\(N_{t}=1:N\)}{
%           \State $ \textbf{J}= fft(\textbf{L}) $
%           \State $ \textbf{J}( abs(\textbf{J}) < 10^{-3})=0 $
%           \State $ \textbf{L'}=ifft(\textbf{J}) $
%           \State $ \textbf{X}(:,N_{t})= reshape(\textbf{L'},[M \ \times \ N,1]) $}
%         \EndFor 
% \end{algorithm}

\section{Numerical Results}
\label{numres} 
We implement our method (ADMD) on three synthetic dataset: double gyre, hidden dynamics, and turbulent channel flow at friction velocity Reynolds number of 1000 and then compare its performance with that of the standard DMD. All the results are produced using Python 3.5 (on an Intel CORE i5 processor with 8 GB 1600 MHz DDR3 memory). Also, it is imperative to note that we would measure uncertainty as,
\begin{equation}
\epsilon = || \textbf{X}_{dmd}(:,t) - \textbf{X}(:,t)||_{2}/ || \textbf{X}(:,t) ||_{2},
\end{equation}
where \(t\) refers to a particular time instant.

\subsection{Double Gyre}
\begin{figure}
    \centering
    \includegraphics[scale=0.7]{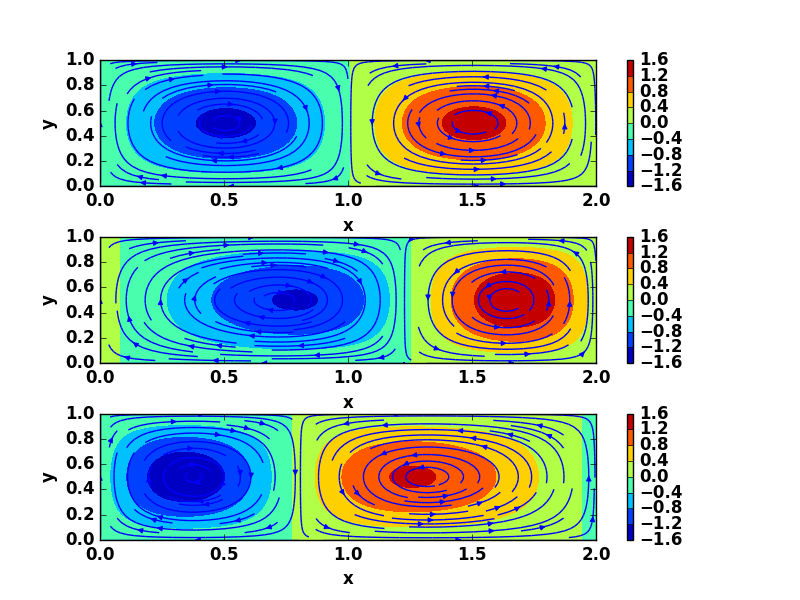}
    \caption{Double Gyre}
    \label{fig:DG}
\end{figure}

\begin{figure}[p]
\centering
\begin{minipage}[c]{.48\textwidth}
  \centering
  \includegraphics[width=1\linewidth]{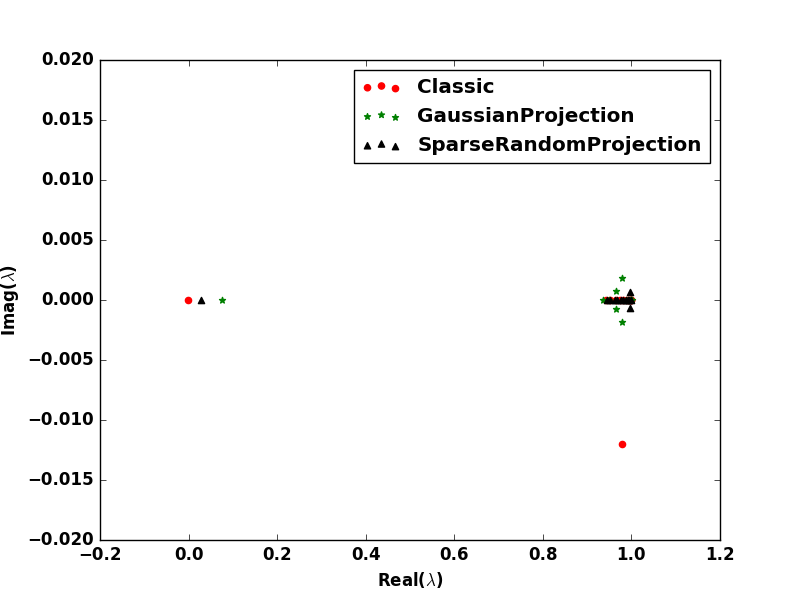}
  \subcaption{(a) Eigenspectrum}
  \label{eigDG}
  
\end{minipage} \hspace{0.1 cm}
\begin{minipage}[c]{.48\textwidth}
  \centering
  \includegraphics[width=1\linewidth]{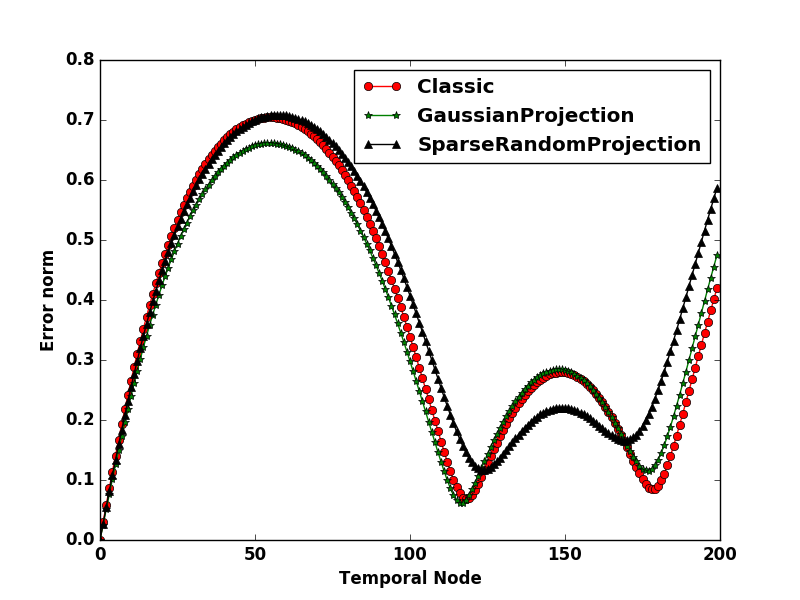}
  \subcaption{(b) Error variation with time}
  \label{errDG}
 
\end{minipage}

\caption{Double Gyre: comparison of performance of standard DMD, ADMD (Gaussian projection assisted), and ADMD (Sparse random projection assisted)}
\end{figure}

A gyre is a system of circulating currents that form due to Coriolis effect. This phenomenon is typically a result of the wind motion through the landmass as the Earth is rotating. A double gyre is an incompressible flow where two counter rotating vortices expand and contract periodically, Figure \ref{fig:DG}. Its model is defined by the stream-function: 
\begin{equation}
 \Psi(x,y,t) = A sin( \pi f(x,t)) sin ( \pi y),
 \end{equation}
 where 
\begin{equation}
f(x,t) = \epsilon sin(\omega t) x^{2} +x -2 \epsilon sin(\omega t) x .
\end{equation}
Note that \((x,y)\) refers to the spatial coordinates and \(t\) the time coordinate.
The horizontal and vertical velocities are derived from the spatial derivatives of the streamfunctions on the domain [0,2] \(\times\) [0,1]: 
\begin{equation}
u= -\frac{\partial \Psi}{\partial y}=\pi A sin(\pi f(x)) cos(\pi y),
\end{equation}
\begin{equation}
v=\frac{\partial \Psi}{\partial x}= \pi A cos (\pi f(x) ) sin( \pi y) \frac{df}{dx}.
\end{equation}
Parameters used are \(A= 0.1\), \(\omega= 2 \pi/10\), \(\epsilon=0.25\). The vorticity, computed from the spatial derivatives of the velocities, will read as
\begin{equation}
vorticity= \frac{\partial v}{\partial x} -\frac{\partial u}{\partial y}.
\end{equation}
Application of standard DMD, on this dataset of vorticity, uses 65536 spatial nodes and results in a reliable model, Table \ref{q1DG}. ADMD identifies the data as of low order and utilizes less than \(1 \%\) of the data to find the dominant features and build a reasonable model which is as good as the one from the standard DMD.\\ \\   
Although, introduction of time-delay coordinates has little or almost no impact on the accuracy of the model, delay embeddings are able to detect a lot more hidden features. This is illustrated by the increase in the rank as delay coordinates are increased, Table \ref{q3DG}. \\ \\
The eigenvalues from ADMD agree with that of standard DMD to a reasonable extent, Figure \ref{eigDG}. One eigenvalue for each method lies close to the origin and the rest tend to move towards the right half plane. Moreover, the error from both DMD and ADMD remain as low as around \(10^{-1}\), Figure \ref{errDG}. The spectrum and the error plot shown apply to the case with no delay coordinates.\\ \\
%The eigenvalues from ADMD agree with that of standard DMD to a reasonable extent, Figure \ref{eigDG}. Most of the eigenvalues in the standard DMD model and the ADMD via 
\begin{table}[h!]
\caption{Double Gyre - no delay coordinates, \(q=1\)}
\centering
 \begin{tabular}{||c c c||} 
 \hline
 DMD variant & Measurements Used & Rank\\ [0.5ex] 
 \hline\hline
 Standard & 65536 & 13\\ 
 Gaussian projection based & 13 & 13 \\
 Sparse random projection based & 20 & 13\\ [1ex] 
 \hline
 \end{tabular}
 \label{q1DG}
\end{table}

\begin{table}[h!]
\caption{Double Gyre - delay coordinates used, \(q=3\)}
\centering
 \begin{tabular}{||c c c||} 
 \hline
 DMD variant & Measurements Used & Rank\\ [0.5ex] 
 \hline\hline
 Standard & 65536 & 24\\ 
 Gaussian projection based & 13 & 26\\
 Sparse random projection based & 20 & 28\\ [1ex] 
 \hline
 \end{tabular}
 \label{q3DG}
\end{table}

%---------------------------------------------
%-------------------------------------------------
\begin{figure}
    \centering
    \includegraphics[scale=0.6]{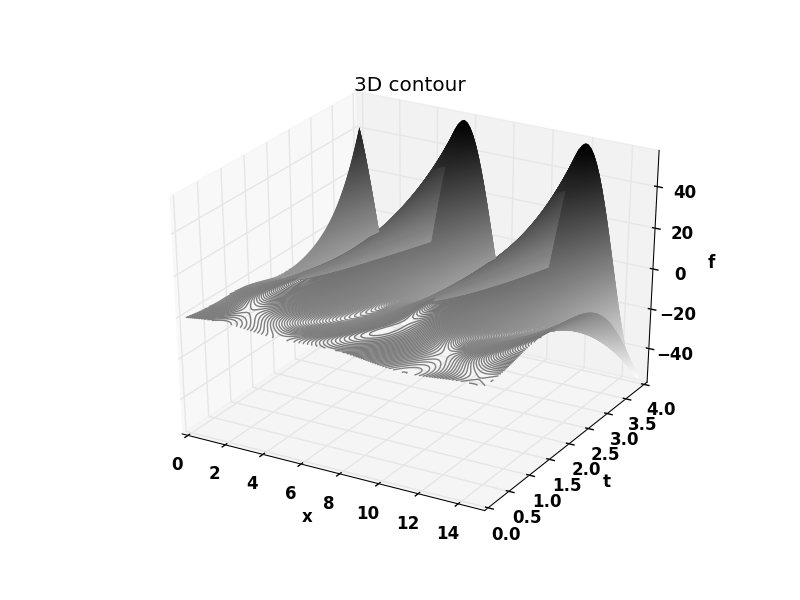}
    \caption{Hidden Dynamics}
    \label{fig:HD}
\end{figure}

\begin{figure}[p]
\centering
\begin{minipage}[c]{.48\textwidth}
  \centering
  \includegraphics[width=1\linewidth]{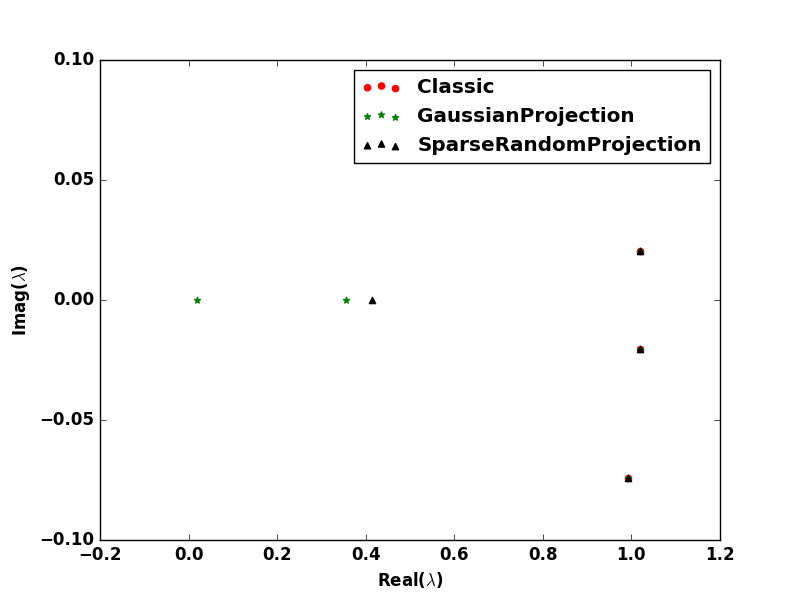}
  \subcaption{(a) Eigenspectrum}
  \label{eigHD}
  
\end{minipage} \hspace{0.1 cm}
\begin{minipage}[c]{.48\textwidth}
  \centering
  \includegraphics[width=1\linewidth]{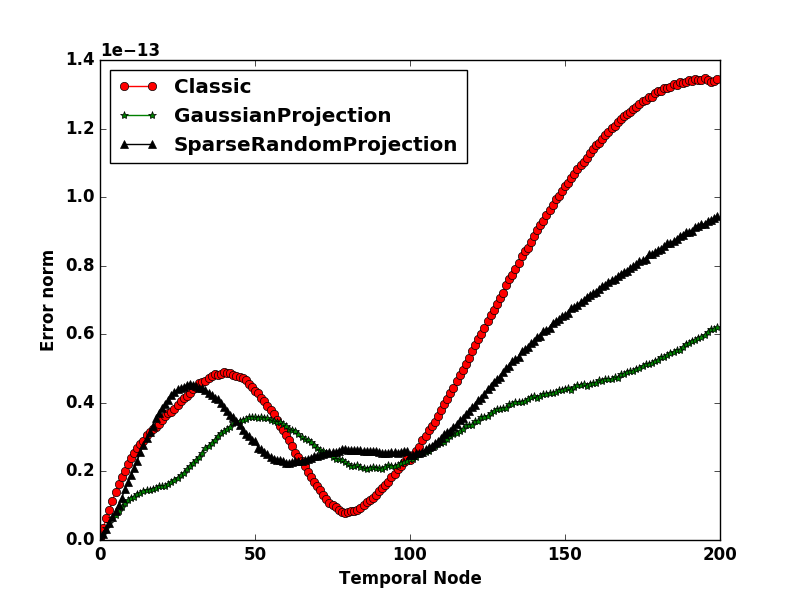}
  \subcaption{(b)Error variation with time}
  \label{errHD}
 
\end{minipage}
\caption{Hidden Dynamics: Comparison of performance of standard DMD, ADMD (Gaussian projection assisted), and ADMD (Sparse random projection assisted)}
\end{figure}

%-------------------------------------------------------
\begin{table}[h!]
\caption{Hidden Dynamics - no delay coordinates, \(q=1\)}
\centering
 \begin{tabular}{||c c c||} 
 \hline
 DMD variant & Measurements Used & Rank\\ [0.5ex] 
 \hline\hline
 Standard & 100 & 4\\ 
 Gaussian projection based & 10 & 6\\
 Sparse random projection based & 10 & 4\\ [1ex] 
 \hline
 \end{tabular}
 \label{q1HD}
\end{table}
\begin{table}[h!]
\caption{Hidden Dynamics - delay coordinates used, \(q=3\)}
\centering
 \begin{tabular}{||c c c||} 
 \hline
 DMD variant & Measurements Used & Rank\\ [0.5ex] 
 \hline\hline
 Standard & 100 & 4\\ 
 Gaussian projection based & 10 & 7\\
 Sparse random projection based & 10 & 4\\ [1ex] 
 \hline
 \end{tabular}
 \label{q2HD}
\end{table}
%------------------------------------------------
\subsection{Hidden Dynamics}
In this example, we consider a dataset produced from the superposition of two sinusoidal functions, one of which sees increasing amplitude and other decreasing amplitude, Figure \ref{fig:HD}. It is defined as,
$$ f(x,t) = sin(k_{1} x - \omega_{1} t) e^{\gamma_{1} t} + sin(k_{2} x - \omega_{2} t) e^{\gamma_{2} t}, $$
where \(k_{1}= 1\), \(\omega_{1}= 1\), \(\gamma_{1}= 1\), \(k_{2}= 0.4\), \(\omega_{2}= 3.7\), and \(\gamma_{2}= -0.2\) for \(0 \le x \le 15\) and \(0 \le t \le 4\). \\ \\
In this problem, 100 spatial nodes are used and DMD uses the complete dataset to create a model. ADMD generates a faithful model (much similar to the model from DMD) using just \(10 \%\) of the data.\\ \\
Time delay coordinates does not have significant impact on the rank of the data, evident from the information in Table \ref{q1HD} and Table \ref{q2HD}. It is worth noting that Gaussian projection has some contribution in determining the hidden modes.\\ \\
The eigenspectrum of DMD and ADMD (with two different projections) are shown in Figure \ref{eigHD}. The eigenvalues from DMD and ADMD (Sparse random projection assisted) agree well, whereas Gaussian projection based ADMD extracts two more eigenvalues with important hidden features: one is \(0.0033 + 0i\) (a background/slow mode) and the other is \(0.136 + 0i\) (a fast mode). The error in the standard DMD model and the ADMD models remains very low (around \(10^{-13}\)) and comparable to each other, Figure \ref{errHD}.

%-------------------------------------------------

\begin{figure}
    \centering
    \includegraphics[scale=0.7]{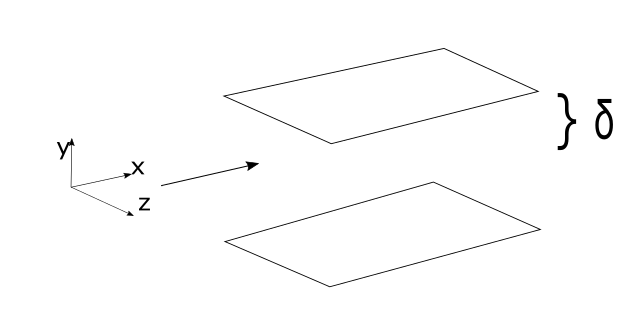}
    \caption{Turbulent Channel Flow}
    \label{fig:tcf}
\end{figure}
  
%--------------------
\begin{table}[h!]
\caption{Turbulent Channel Flow - no delay coordinates, \(q=1\)}
\centering
 \begin{tabular}{||c c||} 
 \hline
 DMD variant & Rank\\ [0.5ex] 
 \hline\hline
 Standard & 2500\\ 
 ADMD via DFT & 2500\\
 ADMD via inexact ALM & 1141\\ [1ex] 
 \hline
 \end{tabular}
 \label{q1TCF}
\end{table}
%---------------------------------------------
\begin{figure}[p]
\centering
\begin{minipage}[c]{.48\textwidth}
  \centering
  \includegraphics[width=1\linewidth]{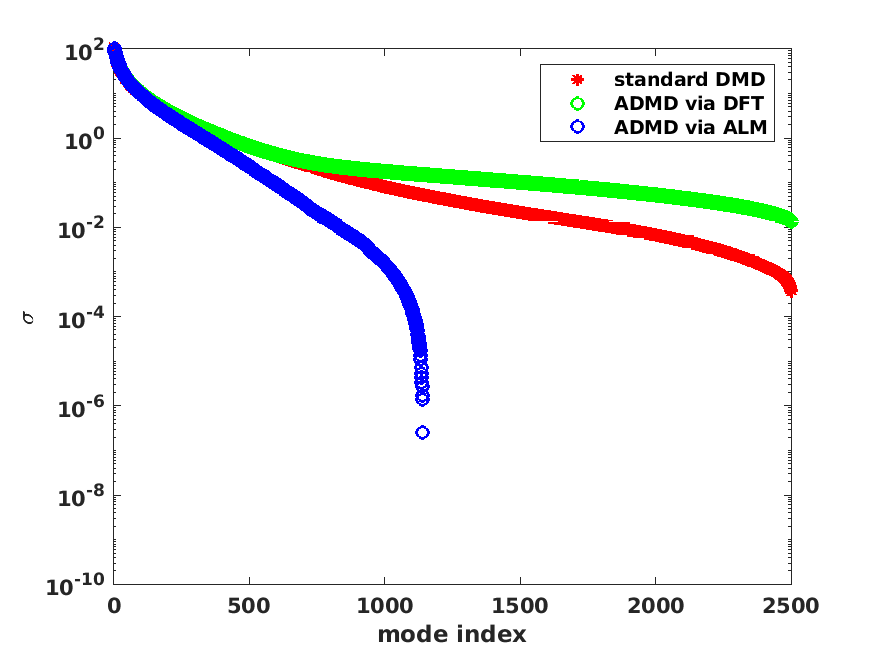}
  \subcaption{(a) Singular values in the data}
  \label{svTCF}
  
\end{minipage} \hspace{0.1 cm}
\begin{minipage}[c]{.48\textwidth}
  \centering
  \includegraphics[width=1\linewidth]{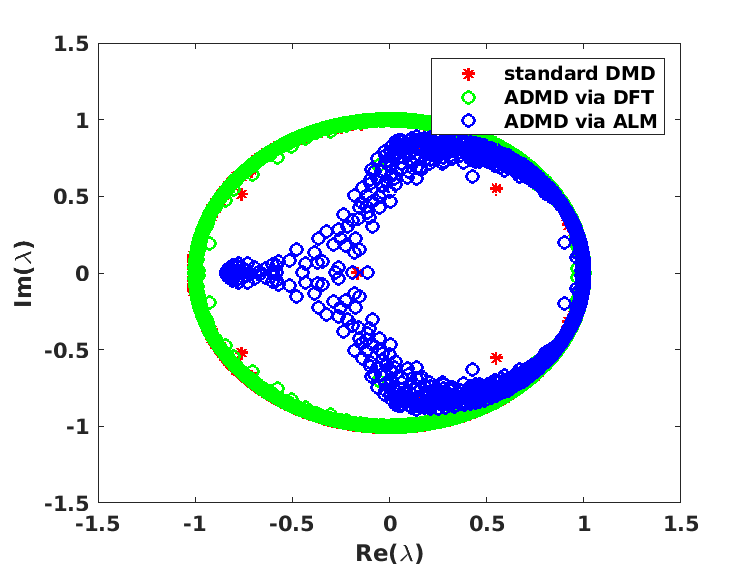}
  \subcaption{(b) Eigenspectrum}
  \label{eigTCF}
 
\end{minipage}

\begin{minipage}[c]{.48\textwidth}
  \centering
  \includegraphics[width=1\linewidth]{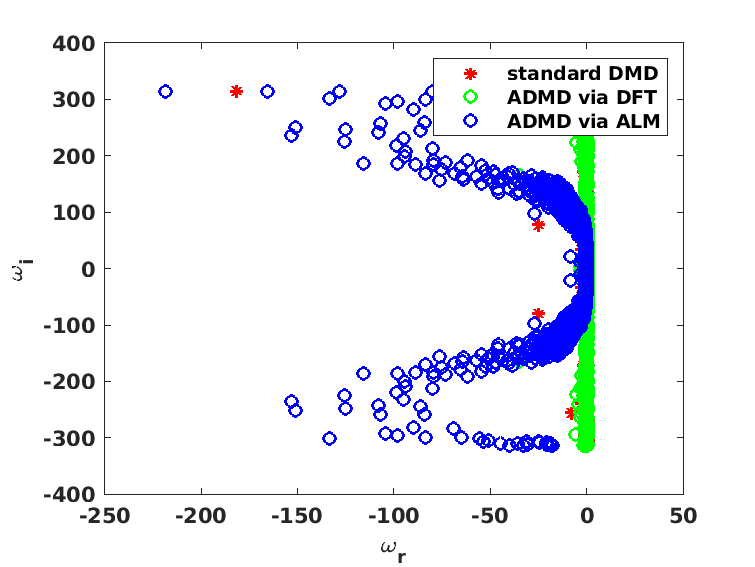}
  \subcaption{(c) Continuous eigenvalue spectrum}
  \label{ceigTCF}
  
\end{minipage}
\begin{minipage}[c]{.48\textwidth}
  \centering
  \includegraphics[width=1\linewidth]{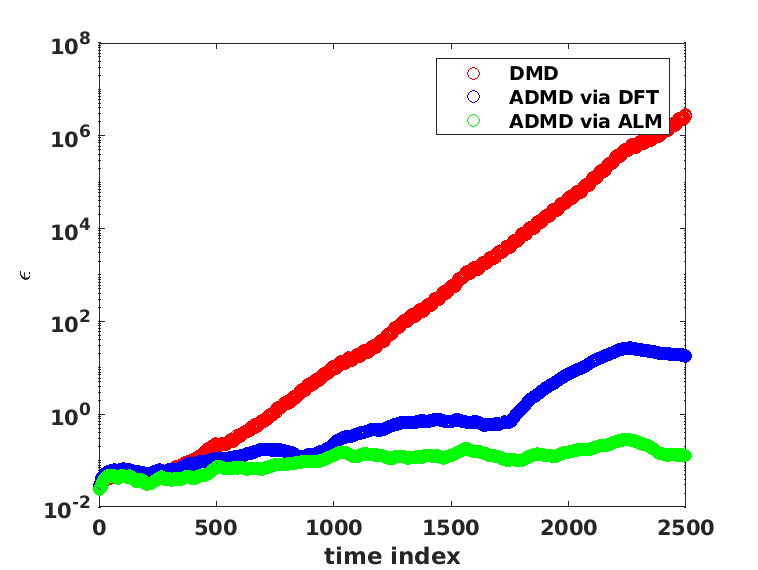}
  \subcaption{(d) Uncertainty with time}
  \label{errTCF}
 
\end{minipage}

\caption{Comparison of DMD and ADMD on Turbulent Channel Flow Data}
\end{figure}  
 %--------------------------------------------------

\subsection{Turbulent Channel Flow}
The data for turbulent channel flow at friction velocity Reynolds number of 1000 is obtained from the John Hopkins University Turbulence Database, \cite{kim1987turbulence}. The geometry with periodic boundary condition is shown in Figure \ref{fig:tcf}. 
Direct numerical simulation is run based on velocity-vorticity formulation. The governing equations are solved using Fourier modes for the longitudinal (\(x\)) and transverse axis (\(z\)) and B-splines collocation method in the wall normal (\(y\)) direction. The flow is started by a bulk channel mean velocity of 1 \(m/s\) until the stationarity is reached and is observed till the fluid moves across the complete domain of interest. This simulation yields field plots of velocity and pressure over time. We use the \(x\)-velocity data from \(N_{t} = 1\) to \(N_{t} = 2500\) to appraise ADMD compared to DMD.\\ \\
We apply the standard DMD and ADMD via discrete Fourier transform and ADMD via inexact augmented Lagrange multiplier method since the dataset is high-dimensional. The singular values from standard DMD are much closer to the ones from the ADMD via DFT, whereas ADMD via inexact ALM finds out a set of quickly decaying singular values, Figure \ref{svTCF}. The rank of the dataset is not much sensitive to DFT, but is greatly reduced by inexact ALM method, Table \ref{q1TCF}. \\ \\
The discrete eigenvalues from standard DMD and ADMD via DFT lie on the unit circle, whereas almost half of the discrete eigenvalues from the ADMD via inexact ALM fall inside the unit disc, Figure \ref{eigTCF}. Most of the continuous eigenvalues from the standard DMD and ADMD via DFT stay on the \(Re(\omega) = 0\) line. On the other hand, the ones from the ADMD via inexact ALM remain on the left half of the plane, Figure \ref{ceigTCF}. This implies that the model from the ADMD via inexact ALM would be more reliable than that from standard DMD and ADMD via DFT. This is confirmed as we plot the error over time for these methods, Figure \ref{errTCF}. The error in the standard DMD model grows faster with time than that in the model from ADMD via DFT. ADMD via inexact ALM turns out to be superior among these three methods since the uncertainty in its model is even lower than that in the model from ADMD via DFT.    
%\section{Declaration of Competing Interest}
%The authors declare that they have no known competing financial interest or personal relationship that can influence the work done in this paper.

\section{Conclusion and Future Work}
\label{cfw}
In this paper, we devised a new version of DMD -- Adaptive Dynamic Mode Decomposition and test its efficiency on three synthetic dataset. ADMD has the capability to understand the character of the data and then accordingly find a faithful model. For low-rank data, ADMD uses projection methods and time-delay coordinates based on the need. In case of high-dimensional data, ADMD leverages discrete Fourier transform and optimizers like inexact augmented Lagrange multiplier method to find a low rank version of the data and then builds an approximate model. The first numerical test was on Double Gyre (a low-rank data-set), where ADMD exploits just a small fraction of the data to make a model as good as the standard DMD model. In the second example, ADMD not only builds a reliable model from just a few measurements, it can also extract hidden features via Gaussian projection of the original dataset. The last numerical test was on a turbulent channel flow at friction velocity Reynolds number of 1000. We see that ADMD generates a model that outperforms the DMD model. In this case, the model from ADMD via inexact augmented Lagrange multiplier method is much more stable model than the one from ADMD via discrete Fourier transform. Overall, ADMD can adapt to data of any nature and produce a model that is better than or as good as the standard DMD model. In future, we would like to make ADMD more robust based on the type of the data and also test its implementation on unsteady, complex flows like screeching jet.
\label{conclu}

\bibliographystyle{IEEEtran}     
\bibliography{main}

\end{document}